\documentclass{emulateapj}
\usepackage{apjfonts}
\newcommand{\gcc}{\ \mathrm{g\ cm^{-3} }}

\newcommand{\cms}{\ \mathrm{cm \ s^{-1}}}

\newcommand{\cm}{\ \mathrm{cm}}

\newcommand{\nuclei}[2]{\ensuremath{\mathrm{^{#1}#2}}}

\newcommand{\carbon}{\nuclei{12}{C}}
\newcommand{\oxygen}{\nuclei{16}{O}}

\newcommand{\nickel}{\nuclei{56}{Ni}}

\newcommand{\pder}[2]{\ensuremath{\frac{\partial #1}{\partial #2}}}

\shorttitle{Flame Evolution During a Type Ia}
\shortauthors{Townsley et al.}

\submitted{Accepted to the Astrophysical Journal}
\begin{document} 

\title{Flame Evolution During Type Ia Supernovae and the Deflagration Phase
in the Gravitationally Confined Detonation Scenario}

\author{
D.~M.~Townsley\altaffilmark{1,2},
A.~C.~Calder\altaffilmark{3,1,$\dagger$},
S.~M.~Asida\altaffilmark{4},
I.~R.~Seitenzahl\altaffilmark{2,5},
F.~Peng\altaffilmark{1,2,3,$\ddagger$},
N.~Vladimirova\altaffilmark{3},
D.~Q.~Lamb\altaffilmark{3,1,5},
J.~W.~Truran\altaffilmark{1,2,3,5,6}
}

\altaffiltext{1}{Department of Astronomy \& Astrophysics,
                 The University of Chicago,
                 Chicago, IL  60637}
\altaffiltext{2}{Joint Institute for Nuclear Astrophysics,
                 The University of Chicago,
                 Chicago, IL  60637}
\altaffiltext{3}{Center for Astrophysical Thermonuclear Flashes,
                 The University of Chicago,
                 Chicago, IL  60637}
\altaffiltext{4}{Racah Institute of Physics,
                 Hebrew University,
		 Jerusalem 91904, Israel}
\altaffiltext{5}{Enrico Fermi Institute,
                 The University of Chicago,
                 Chicago, IL  60637}
\altaffiltext{6}{Argonne National Laboratory,
                 Argonne, IL 60439}
\altaffiltext{$\dagger$}{now at 
                 Department of Physics and Astronomy,
                 SUNY, Stony Brook,
                 Stony Brook, NY  11794-3800}
\altaffiltext{$\ddagger$}{now at 
                 Theoretical Astrophysics,
                 California Institute of Technology,
                 Pasadena, CA  91125}

\begin{abstract}

We develop an improved method for tracking the nuclear flame during the
deflagration phase of a Type Ia supernova, and apply it to study the
variation in outcomes expected from the gravitationally confined detonation
(GCD) paradigm.  A simplified 3-stage burning model and a non-static ash
state are integrated with an artificially thickened
advection-diffusion-reaction (ADR) flame front in order to provide an
accurate but highly efficient representation of the energy release and
electron capture in and after the unresolvable flame.  We demonstrate that
both our ADR and energy release methods do not generate significant acoustic
noise, as has been a problem with previous ADR-based schemes.  We proceed to
model aspects of the deflagration, particularly the role of buoyancy of the
hot ash, and find that our methods are reasonably well-behaved with respect
to numerical resolution.  We show that if a detonation occurs in material
swept up by the material ejected by the first rising bubble but
gravitationally confined to the white dwarf (WD) surface (the GCD paradigm), the density
structure
of the WD at detonation is systematically correlated with the distance of the
deflagration ignition point from the center of the star.  Coupled to a suitably stochastic
ignition process, this correlation may provide a plausible explanation for the variety of
nickel masses seen in Type Ia Supernovae.

\end{abstract}

\section{Introduction}

It is widely believed that in a type Ia supernova explosion, a WD near the
Chandrasekhar limiting mass is disrupted by a thermonuclear runaway in its
interior,  and more precisely that a subsonic deflagration must precede 
any detonation (see \citealt{hillebrandt+00} and references
therein).  The current leading paradigms for how the deflagration of a WD
takes place, and how this leads to astrophysical properties that match
observations, are generally termed (1) pure deflagration, (2) deflagration
detonation transition (DDT), (3) pulsational detonation, and (4)
gravitationally confined detonation (GCD,~\citealt{plewa+04}).  In all but
the first of these, a subsonic deflagration phase expands the WD, lowering
its density, and a subsequent supersonic detonation then incinerates the
remainder of the star.  Among the remaining three, the process that is
proposed to ignite the detonation is very different, though it is crucial to
determine how much expansion can occur prior to the detonation in order to
predict the variation of nickel mass and therefore brightness among the
observed Type Ia's.

A primary purpose of this work is to set out a numerically efficient method
for modeling the nuclear energy release in the flame front that
propagates via heat diffusion during the deflagration stage.  This
formalism will be used for studying a variety of features of all of the above
paradigms in future work.  Nucleosynthesis of species produced as a result of
electron captures provides a very important observational constraint on
supernova models, especially the pure-deflagration scenario.  For this
reason, and because methods are available in the literature \citep{gamezo+05},
we have included electron capture and neutrino emission in the energetic
treatment to capture its effect on the hydrodynamics.  Description of this
method incorporates our previous work \citep{calder+07} detailing the nuclear
processing of \carbon\ and \oxygen\ by a flame front and the evolution of
the resulting ash.  A method for integrating this simplified 3-stage energy
release with an artificially broadened flame is described in Section
\ref{sec:model}.  The acoustic properties of this method are discussed in
Section \ref{sec:noise}, where it is shown that the front emits very little
acoustic noise.  This is important for reducing spurious seeding of the
strong hydrodynamic instabilities present during the deflagration phase.

We have chosen in this work to initially pursue simulations of the 
deflagration phase in GCD because it
provides a more direct demonstration of the buoyancy character of the flame
bubble and current work on this mechanism
\citep{plewa+04,calder-asym+04,plewa+07,roepke-astroph+06} can benefit from a concise
parameter study.  In this mechanism, the strong eruption of a rising flame
bubble through the surface creates a wave of material traveling over the
surface that collides at the point opposite breakout, compressing
and heating unburnt surface material until detonation conditions are reached.  We therefore proceed in
section \ref{sec:progression} to discuss our setup for simulating the
deflagration of the star, in which our principle hypothesis is that the first
flame ignition point is rare and therefore the deflagration phase is
dominated by a single flame bubble.  Some perspective is given with respect
the conditions expected to be present in the WD core at this time, and we
describe the progression of the burning in the simulation, including a survey
of the effects of simulation resolution.  Finally, in section
\ref{sec:discussion} we present the results of simulations in which the ignition
point of the flame is placed at various distances from the center of the
star.  We find that the density of the star at the time when the GCD
mechanism predicts an ignition of the detonation, and thus the mass that will
be processed to Fe group elements, is well correlated with the offset of the
initial ignition point.  This parameter study also serves as a touch-point for
future larger-scale simulations of this mechanism in three dimensions
\citep{jordan+07}, which are
essential for judging its viability \citep{roepke-astroph+06}.  We summarize
and make some concluding remarks in section \ref{sec:conclusion}.


\section{Burning Model for a Carbon Oxygen White Dwarf}
\label{sec:model}

There are two fairly different methods of flame-front tracking used in 
contemporary studies of 
WD deflagrations.  Use of a front-tracking method is necessary
because the physical thickness of the flame front is unresolvable in any
full-star scale simulation, with the carbon consumption stage being $10^{-4}$ to
$10^3$ cm thick for the density range important in the star
\citep{calder+07}.  The method presented here is based upon
propagating a reaction progress variable with an advection-reaction-diffusion
(ADR) equation, and can be thought of as an artificially thickened flame,
because the real flame is also based on reaction-diffusion on much smaller scales.
This type of method has been used in many previous simulations of the WD
deflagration, both in full star simulations (e.g.
\citealt{gamezo+03,calder-asym+04,plewa+07})
and to study the effect of the Rayleigh-Taylor (R-T)
instability on a propagating flame front \citep{khokhlov+95,zhang+06}.  Our flame
propagation is based heavily on this work, and we have made several
refinements to the method that we will describe in detail below.  The other
widely used method utilizes the level set technique
\citep{smiljanovski+97,reinecke+99,MPA03} and performs an interface
reconstruction in each cell based upon the value of a smooth field defined on
the grid and propagated with an advection equation acting in addition to the
hydrodynamics. See \citet{roepke-astroph+06} and \citet{schmidt+06} for
recent deflagration simulations using this method.

It should be emphasized that the implementation of the flame propagation is
far from the only difference between these approaches, and there is
considerable latitude even within one of the front-tracking methods.  In
addition to the front-tracking itself two other issues are important.  First,
the energy release of the nuclear burning must be treated, and this is
typically done in some simplified way for computational efficiency.  For example,
a prominent difference between the method presented here and that commonly
used with level-set is that we include electron captures in the post-flame
material within our treatment.  The second important additional component is
what measure is taken to prevent the breakdown of the flame tracking method
when R-T, and possibly secondary instability in the induced flow, is strong
enough to drive flame surface perturbations on a sub-grid scale.  Both
methods fail in this limit because the scalar field being used to propagate
the flame is distorted by advection due to strong turbulence.
Generally this has been overcome by increasing the flame speed enough
to polish out grid-scale disorder in the flow field.  This can, however, be
phrased in terms of simple \citep{khokhlov+95} or complex \citep{schmidt+06}
laws intended to mimic the enhanced flame surface area produced
by unresolved structure in the flame.  We will leave further discussion to
separate work, but awareness of this difference is important for comparing
results of the two methods.

\subsection{ADR Flame-front Model}

Generally, an ADR scheme characterizes the location of a flame front using a reaction
progress variable, $\phi$, which increases monotonically across the front
from 0 (fuel) to 1 (ash).
Evolution of this progress variable is accomplished via an
advection-diffusion-reaction equation of the form
\begin{equation}
\pder{\phi}{t}+\vec v\cdot\nabla\phi = \kappa\nabla^2 \phi + \frac{1}{\tau}R(\phi)\ ,
\end{equation}
where $\vec v$ is the local fluid velocity, and the reaction term, $R(\phi)$,
timescale, $\tau$, and the diffusion constant, $\kappa$, are chosen so that
the front propagates at the desired speed.
\citet{vladimirova+05} showed that the step-function reaction rate widely in
use led to a substantial amount of unwanted acoustic noise.  They
studied a suitable alternative, the Kolmogorov Petrovski Piskunov (KPP)
reaction term which has an extensive history in the study of
reaction-diffusion equations.  In the KPP model the reaction term is given by
\begin{equation}
R(\phi) = \frac{1}{4} \phi(1-\phi)\ .
\end{equation}
The symmetric and low-order character of this function gives it very nice
numerical properties, leading to amazingly little acoustic noise.  Following
\citet{vladimirova+05}, we adopt $\kappa\equiv sb\Delta x/16$ and $\tau
\equiv b\Delta x/16s$, where $\Delta x$ is the grid spacing, $s$ is the
desired propagation speed, and $b$ sets the desired front width scaled to
represent approximately the number of zones.

The KPP reaction term, however, has two serious drawbacks.  Formally, the
flame speed is only single valued for initial conditions that are precisely
zero (and stay that way) outside the burned region
\citep{Xin00}, which cannot really be effected in a hydrodynamics simulation.
This can lead to an unbounded increase of the propagation speed, which is
precisely the property we wish to have under good control.  Secondly, the
progress variable $\phi$ takes an infinite amount of time to actually reach
1 (complete consumption of fuel).  While not a fatal flaw like the flame speed 
problem, this is a problem
for our simulations in which we would like to have a localized flame front so
that fully-burned ash can be treated as pure NSE material.

Both of these drawbacks can be ameliorated by a slight modification of the
reaction term \citep{asida+07} to
\begin{equation}
\label{eq:skpp}
R(\phi) = \frac{f}{4}(\phi-\epsilon_0)(1-\phi+\epsilon_1)\ ,
\end{equation}
where $0<\epsilon_0, \epsilon_1\ll 1$ and $f$ is an additional factor that
depends on $\epsilon_0$ and $\epsilon_1$ and the flame width so that the
flame speed is preserved with the same constants as for KPP above.  This ``sharpened''
KPP (sKPP) has truncated tails in both directions (thus being sharpened),
making the flame front fully localized, and is a bi-stable reaction rate and
thus gives a unique flame speed \citep{Xin00}.  The price paid is that since
$R(\phi)=0$ for $\phi \le 0$ and $\phi \ge 1$, (\ref{eq:skpp}) is
discontinuous, adding some noise to the solution.  Since the suppression of
the tails is stronger for higher $\epsilon_0$ and $\epsilon_1$, we adjusted
$\epsilon_0$ and $\epsilon_1$ so that for a particular flame width we could
meet our noise goals. The parameter values used in the simulations 
presented in this work were $\epsilon_0 = \epsilon_1 = 10^{-3}$, $f = 1.309$, 
and $b = 3.2 $.  The noise properties of these choices 
are discussed in section \ref{sec:noise}.

Diffusive flames are known to be subject to a curvature effect that affects
the flame speed when the radius of curvature is similar to the flame
thickness, a frequent circumstance with modestly-resolved flame front
structure.  In testing, the curvature effect of the step-function reaction
rate proved surprisingly strong, likely due to the exponential ``nose'' that
the flame front possesses \citep{vladimirova+05}.  Both KPP and our sKPP show
significantly better curvature properties.  Due to the necessary discussion
of background and the size of the study supporting this conclusion, 
this topic will be discussed in detail separately
\citep{asida+07}.

\subsection{Brief Review of Carbon Flame Nuclear Burning in White Dwarfs}

In previous work \citep{calder+07}, we performed a detailed study of the
processing that occurs in the nuclear flame front and
the ashes it leaves.  It was shown that, as discussed previously 
\citep{khokhlov+83,Khokhlov91+dd}, the nuclear burning proceeds in roughly 3
stages: consumption of \carbon\ is followed by consumption of \oxygen\ on a
slower timescale, which is in turn followed by conversion of the
resulting Si group nuclides to Fe group.  Most of the energy release takes
place in the $\carbon$ and $\oxygen$ consumption steps, and at high
densities the resulting material contains a significant fraction of light
nuclei ($\alpha$, $p$, $n$) and is in an active equilibrium in which
continuously occurring captures of the light nuclei are balanced by their
creation via photodisintegration.  Initially, the heavy nuclei are
predominantly Si group, this is termed nuclear statistical quasi-equilibrium
(NSQE), which upon conversion of these to Fe group becomes nuclear
statistical equilibrium (NSE).  Each of these states is reached on a
progressively longer timescale, and the importance of distinguishing the
last lies in the disparity in electron capture rates between Si and Fe
group based equilibria.
The energy released by our scheme at a given density has
been directly verified within a few percent against those tabulated in
\citet{calder+07} and against an additional direct NSE solution.

Our methods build heavily on those of \citet{gamezo+05} and
\citet{Khokhlov91+dd} (see also \citealt{khokhlov+00}), which is used
throughout their family of recent deflagration calculations
\citep{gamezo+03,gamezo+04,gamezo+05}.  The principal differences, other than
the use of the sKPP reaction term, are that we use the predicted binding energy,
$q_f$ (see definitions below), of the \emph{final} NSE state rather than using $q_{\rm
nse}(\rho,T,Y_e)$ with the current density and temperature, $\rho$ and $T$,
and we separately track \carbon\ and \oxygen\ consumption.  These are
described in detail below.  Finally, the method presented here is entirely
different from that used by \citet{plewa+07}, which effectively ``freezes'' the
NSE at the state produced in the flame front, neglecting the additional
energy release as the light nuclei are recaptured.

\subsection{A Quiet Three Stage, Reactive Products Flame Front}

In incompressible simulations, the progress variable in an ADR front tracking
scheme is typically used to parameterize the density or density decrement.
An analog in compressible simulations is to release energy in proportion to
$\phi$.  This simple idea becomes somewhat complicated in a situation like
the WD, where the burning (and therefore energy release) occurs in multiple
stages whose progress time scales vary by orders of magnitude during the
simulation.  A further complication is created by the dynamic NSE state of
the ash, such that the energy release depends on the physical conditions
(density) under which the flame is evolving.

The ethos we have implemented here is that processes that occur on scales
that are unresolved by the artificially thickened flame should have their
energy release counted towards the \emph{overall} energy that is smoothly released by
the progression of the flame variable $\phi$.  This approach is accomplished by
defining additional progress variables that follow the ADR variable $\phi$
and that govern the energy release.  Such a complex scheme is necessary for
the nuclear burning in the WD because, as shown in \citet{calder+07}, the
conversion of Si to Fe group that occurs over centimeters near the core,
occurs over kilometers in the outer portion of the star.
In previous work \citep{calder+07}, we presented a method for integrating
energy release with an ADR flame. Those prescriptions were an early version
of what is presented here, and are superseded by the method presented below.
Note in particular that the functional meaning of $\phi_2$ and $\phi_3$ have
changed somewhat because the ash state is able to evolve regardless of the
value of the progress variables.  Also, the use of surrogate nuclei described
in that work has been abandoned in favor of the direct use of scalars
described below.

We define three progress variables, which represent irreversible processes.
These three variables start at 0 in the unburned fuel and progress
toward 1, representing
\begin{eqnarray*}
\phi_1 &\quad&\text{Carbon consumption, conversion of C to Si group}\\
\phi_2 &\quad&\text{Oxygen consumption, conversion of O to Si group}\\
\phi_3 &\quad&\text{Conversion of Si group to Fe group.}
\end{eqnarray*}
We keep strictly $\phi_1\ge \phi_2 \ge\phi_3$, but all three are allowed to have
values other than zero and 1 in the same cell.  It is most useful to think of
material in a cell as being made up of mass fractions of $1-\phi_1$ of
unburned fuel, $\phi_1-\phi_2$ of partially burned (no carbon) fuel, $\phi_2$
of NSQE material of which $\phi_3$ has had its Si group elements consumed.
As shown by \citet{calder+07}, given sufficient resolution, all these stages
are, in fact, discernible as fairly well separated transitions.  However, with
an artificial flame, a real transition from fuel to final ash that occurs in
less than one grid spacing must be spread out over several.

We now describe how these auxiliary progress variables track the flame
progress.  In our case the evolution of $\phi_1$ is set directly by the
artificial flame formalism described above, $\phi_1 \equiv \phi$.  Thus the
noise properties of the artificial flame itself are inherited by the energy
release scheme.  The connection between the energy release and the ADR flame
tracking comes entirely through this equality, and so coupling the following
energy release methodology to other available front-tracking methods appears
quite practicable.

We evolve a number of scalars which, in the absence of sources, satisfy
a continuity equation,
\begin{equation}
\label{eq:massscalar}
\pder{Q\rho}{t} +\nabla\cdot(Q\rho \vec v) = 0\ ,
\end{equation}
where $Q$ is the scalar under consideration.  The flame variable $\phi$
above is one such scalar, and our additional progress variables are also
treated as such.  The other scalars we utilize directly represent
physical properties of the flow; they are the number of electrons, the number
of ions and nuclear binding energy per unit mass or baryon, respectively:
\begin{eqnarray}
\label{eq:ye}
Y_e &=& \sum_i \frac{Z_i}{A_i}X_i\ ,\\
Y_{\rm ion} &=& \sum_i \frac{1}{A_i} X_i = \frac{1}{\bar A}\ ,\\
\label{eq:qbar}
\bar q & =& \sum_i \frac{E_{b,i}}{A_i} X_i\ ,
\end{eqnarray}
where $i$ runs over all nuclides, $Z_i$ is the nuclear charge, $A_i$ is the
atomic mass number (number of baryons), and
$E_{b,i} = (Z_i m_p - N_im_n-m_i)c^2$
is the nuclear binding
energy where $Z_i$, $m_p$, $N_i$, and $m_n$ are the number and rest mass of
protons and neutrons respectively, so that positive is more bound.
The mass
fractions $X_i$ are \emph{not} treated in our simulation, and are used here only to
define these properties, though (\ref{eq:ye})-(\ref{eq:qbar}) satisfy
(\ref{eq:massscalar}) by virtue of being linear combinations of the mass
fractions, which themselves satisfy (\ref{eq:massscalar}).  Defining our
flame model is then a matter of setting out the source terms for the various
scalars $\phi_1$, $\phi_2$, $\phi_3$, $Y_e$, $Y_{\rm ion}$, and $\bar
q$, and relating these to the energy release.

For a given $\phi_1$ and $\phi_2$ we define, for notational convenience,
\begin{eqnarray}
\label{eq:Xc}
X_{\rm C} &\equiv& (1-\phi_1)X_{\rm C}^0\\
X_{\rm O} &\equiv& (1-\phi_2)(1-X_{\rm C}^0)\\
\label{eq:Xmg}
X_{\rm Mg} &\equiv& (\phi_1-\phi_2)X_{\rm C}^0\ ,
\end{eqnarray}
where $X_{\rm C}^0$ is the initial carbon fraction.  These can serve as
approximate abundances, though the real abundances in these stages have
several additional important species.  Since this notation can be misleading,
we again emphasize that abundances are not being tracked in our simulation,
the material properties used in the EOS are derived directly from $Y_e$ and
$Y_{\rm ion}$, discussed further below.  The remaining mass fraction of
material $\phi_2$ is
considered to be in NSQE or NSE, so that $\phi_2 +X_{\rm C} + X_{\rm O}
+X_{\rm Mg} = 1$.  We define this ``ash'' material to have binding energy
$\bar q_{\rm ash}$ and electron fraction $Y_{e,\rm ash}$ and ion number
$Y_{\rm ion, ash}$ such that,
\begin{equation}
\label{eq:totqbar}
\bar q = \phi_2 \bar q_{\rm ash} + X_{\rm C} q_{\rm C} + X_{\rm O} q_{\rm O}
+ X_{\rm Mg} q_{\rm Mg}
\end{equation}
and similarly for the other quantities.  To again clarify our notation,
$q_{\rm \{C,O,Mg\}}$ are the actual binding energies of $^{12}$C, $^{16}$O
and $^{24}$Mg, being used here to approximate the binding energy of the
intermediate ash state.  The final scalar, $\phi_3$, represents the degree to
which the ash has completed the transition from Si-group to Fe-group heavy
nuclei, and is used to scale the neutronization rate as described below.
Thus after material has expanded and is no longer $\alpha$-rich, $X_{\rm
Si-group}\approx \phi_2-\phi_3$.

From the quantities $\bar q$ and the local internal energy per mass,
$\mathcal E$, it is possible to predict the final burned state if density,
$\rho$, or pressure,
$P$, were held fixed for infinite time and weak interactions (e.g. electron
captures) were forbidden (constant $Y_e$); this gives the NSE state.  Our
equations and formalism for
NSE, which include plasma Coulomb corrections, were described by
\citet{calder+07}.  Using these, the abundances and therefore average binding
energy of the NSE state can be found for a given $\rho$, $T$, and $Y_e$,
resulting in $\bar q_{\rm NSE}(\rho,T, Y_e)$.  $Y_{\rm ion,\rm NSE}$, as well
as the Coulomb coupling parameter $\Gamma={\bar{Z}}^{5/3} e^2(4\pi
n_e/3)^{1/3}/kT$, where e is the electron charge, $n_e$ is the electron
number density and $k$ is Boltzmann's constant, are similarly determined.
For internal energy, $\mathcal E$, we follow the convention of
\citet{timmesswesty2000}, which excludes the rest mass energy of the (matter)
electrons.

The final burned state at a given $\rho$ and $Y_e$ can be found by
solving
\begin{equation}
\label{eq:isochoric}
 \mathcal E -\bar q = \mathcal E(T_f) -\bar q_{\rm NSE}(T_f)
\end{equation}
for $T_f$.  The number of protons, neutrons and electrons is the same in both
states, so that the rest masses cancel in this equation.
We will denote $\bar q_f \equiv \bar q_{\rm NSE}(T_f)$, as the
solution to this or the corresponding isobaric equation below.
For the sake of computational efficiency, this solution is accomplished
via a table lookup in a tabulation of $\bar q_f(\rho,Y_e, \mathcal E-\bar
q)$.  Since the flame is quite subsonic, it is also useful to be able to
predict the NSE final state for the local $P$.  This can be accomplished by
solving, at a particular $P$ and $Y_e$,
\begin{equation}
 \mathcal E-\bar q +\frac{P}{\rho}
= \mathcal E(T_f) -\bar q_{\rm NSE}(T_f) +\frac{P}{\rho_(T_f)},
\end{equation}
or, more naturally,
\begin{equation}
\label{eq:isobaric}
\mathcal H-\bar q = \mathcal H(T_f)-\bar q(T_f),
\end{equation}
where $\mathcal H$ is the enthalpy per unit mass.  This leads to a similar
tabulation of $\bar q_f (P,Y_e, \mathcal H-\bar q)$.  We denote the solution
of (\ref{eq:isochoric}) as isochoric and that of (\ref{eq:isobaric}) as
isobaric.  While the isobaric prediction of the final state must be used
within the thickened flame front, away from the flame front we would like to
use the isochoric result to avoid undue interference with the hydrodynamic
evolution.  This necessitates a hand-off when the material is nearly fully
burned.  We wish the hand-off to occur at a high enough $\phi$ that the
difference between the isochoric and isobaric results are minimized, but we
introduce a small region where the results are explicitly averaged in order
to minimize the noise generated by the hand-off.
Thus, where $\phi_2 > 0.9999$ we use the isochoric estimate, for $0.99
<\phi_2 <0.9999$ we use a linear admixture of isochoric and isobaric, and for
$\phi_2 < 0.99$ we use the isobaric estimate.  From noise tests and behavior
in simulations, these values appear sufficient for the current purposes.

We now have in hand an estimate of the NSE final state $\bar q_f$, and its
temperature $T_f$.  \citet{calder+07} evaluated the timescales for
progression of the burning stages from self-heating calculations, as
functions of $T_f$.  The progress variables are then evolved according
to\footnote{Here $\dot X$ denotes a Lagrangian time derivative, though since
our code is operator-split between the hydrodynamics and source terms, the
implementation is a simple time difference.}
\begin{eqnarray}
\label{eq:phi2dot}
\dot\phi_2 &=& \frac{\phi_1-\phi_2}{\tau_{\rm NSQE}(T_f)}\ ,\\
\label{eq:phi3dot}
\dot\phi_3 &=& \frac{\phi_2-\phi_3}{\tau_{\rm NSE}(T_f)}\ ,
\end{eqnarray}
and the binding energy of the ash material according to
\begin{equation}
\label{eq:qbarashdot}
{\dot{\bar q}}_{\rm ash} = 
\frac{\dot\phi_2}{\phi_2} q_f + \frac{q_f-q_{\rm ash}}{\tau_{\rm NSQE}(T_f)}
\ .
\end{equation}
This is in fact implemented conservatively in the finite difference form 
\begin{equation}
 q_{\rm ash}^{n+1} = \left[\phi_2\left(q_{\rm ash}^{n}+\frac{q_f-q_{\rm ash}}{\tau_{\rm
NSQE}(T_f)}\Delta t\right) + q_f\dot\phi_2\Delta t\right]/(\phi_2+\dot\phi_2\Delta t)\ ,
\end{equation}
where $\Delta t$ is the timestep.
The ion number is treated similarly, according to
\begin{equation}
\dot Y_{\rm ion, ash} = \frac{\dot \phi_2}{\phi_2} Y_{{\rm ion},f}
+\frac{ Y_{{\rm ion},f} - Y_{\rm ion}}{\tau_{NSQE}(T_f)}\ ,
\end{equation}
with a similar finite differencing.  Neutronization (mainly electron
captures) is implemented by applying
\begin{equation}
\dot Y_{e,\rm ash} = \phi_3\dot Y_e(\rho,T_f, Y_e)\ .
\end{equation}
Our calculation of $\dot Y_e$ is described in \citet{calder+07} and utilizes
443 nuclides in the NSE calculation including all
available rates from \citet{langanke+00}.
Finally $\bar q$ is recalculated with the new values of $\phi_1$, $\phi_2$
and $\bar q_{\rm ash}$ using (\ref{eq:totqbar}) and the energy release 
rate is
\begin{equation}
\label{eq:enuc}
\epsilon_{\rm nuc} = \frac{\Delta \bar q}{\Delta t}
-\phi_3[ \dot Y_eN_Ac^2(m_p+m_e-m_n) + \epsilon_\nu]\ .
\end{equation}
Here $\epsilon_\nu$ is the energy loss rate from radiated neutrinos and
antineutrinos, and is calculated similarly to $\dot Y_e$.

Our description here has been fairly algorithmic, for the sake of clarity of
implementation.  It is possible, however, to use eq.
(\ref{eq:phi2dot})-(\ref{eq:qbarashdot}) along with the definitions
(\ref{eq:Xc})-(\ref{eq:Xmg}) and (\ref{eq:totqbar}) and (\ref{eq:enuc}) to
obtain straightforward Lagrangian source terms.

\section{Quantifying acoustic noise from the numerical flame front}
\label{sec:noise}

The noise generated by the model flame may influence the 
outcome of a deflagration simulation by seeding spurious 
fluid instabilities.
Quantifying noise, determining the sources of noise,
and minimizing noise are therefore necessary steps 
in the development of a robust flame model. 
To this end, we performed a suite of simple test
simulations following those of~\citet{vladimirova+05}. The
simulations presented here are for the sKPP flame model with
$\epsilon_0=\epsilon_1 = 10^{-3}$, the highest values for which the RMS
deviation in velocity (see below) was a few$\times10^{-4}$ or lower for
the density range $10^{7}$-$10^9\gcc$.
We note that simulations of model flames
utilizing the ``top hat'' reaction produced considerably
more noise, $\sim 0.1$ or more RMS velocity deviation.

\subsection{Details of Test Simulations}
\label{sec:simdet}

The simulations consisted 
of one-dimensional flames propagating through 40 km of uniform 
density material composed of 50\% \carbon\ and 50\% \oxygen\
by mass.
The simulations had a reflecting boundary condition on the left side
and a zero-gradient outflow boundary condition on the right with
the flame propagating from left to right. The flame was ignited
by setting the left-most 5\% of the domain to conditions expected
for fully burned material, with the transition to unburned fuel 
described by the expected flame profile. This method of ignition
approximated what would have resulted from letting the
flame burn across the ignition region. 

The choice of
boundary conditions allowed material to flow to the right
and off the grid as the flame propagated and more of the 
domain consisted of (expanded) ash.  The densities, sound 
speeds, and sound crossing times of the simulation domain are given in 
Table~\ref{tab:noise}. The simulations were performed on domains of
256, 512, and 1024 zones, corresponding to resolutions of 
15625, 7812, and 3906 $\cm$ respectively. 
The simulations were performed with the FLASH code~\citep{fryxell+00,calder+02},
which explicitly evolves the equations of hydrodynamics. 
In an explicit method such as this, the
time step of a given simulation is limited by the sound crossing time
of the zones.  For these simulations, the maximum time step was determined
by a CFL limit of 0.8, meaning that the time step allowed a sound wave
to cross only eight tenths of the zone with the highest sound speed. 

Acoustic noise may be quantified by considering the magnitude of 
variations in quantities like pressure and velocity. We define the
``RMS deviation'' of a quantity $x$ as 
\begin{equation}
{\mathrm{dev}}_{\mathrm{RMS}}(x) = \sqrt{ \langle x^2 \rangle - \langle x
\rangle^2} / \langle x\rangle\ ,
\end{equation}
where the averages are taken in space at a given time.
In each simulation we calculated the RMS deviation of pressure and 
velocity in both the fuel and the ash, with fuel defined as the
region of the domain with $\phi_1 < 0.0001$ and ash defined as the
region with $\phi_1 > 0.9999$. The RMS deviations presented 
in the figures here are for the fuel.

The densities we considered ranged from 
$\approx 10^7$ to $ 2 \times 10^9 \gcc$. 
Nuclear flame speeds vary extremely with density, from $\approx 5 
\times 10^3$ to $8 \times 10^6 \cms$ for laminar flame speeds at these 
densities~\citep{timmes+92,chamulak+07}. 
Because of the disparity of flame
speeds, we performed some simulations with a fixed flame speed
of $6 \times 10^6 \cms$, allowing flames to propagate
across the simulation domains in similar elapsed time. 
Note the time for the
flame to cross the domain is shorter than the domain size over the flame
speed due to expansion across the flame, which itself depends on density.

\begin{figure}
\plotone{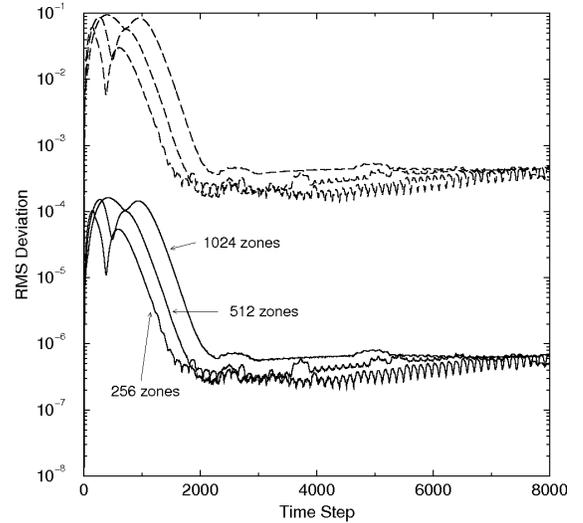}
\caption{\label{fig:r1e9pv}
Plot of RMS deviation in pressure and velocity for simulations of 
propagating model flames 
at a density of $10^9 \gcc$ 
performed
on simulation domains of 256, 512, and 1024 zones.
Shown are the RMS deviations
as functions of the number of time steps, with solid curves indicating
pressure and dashed curves indicating velocity. 
In these, the flame speeds were $3.89 \times 10^6 \cms$, the expected
flame speed for material of this density.
}
\end{figure}

Figure~\ref{fig:r1e9pv} shows the RMS deviation in pressure and velocity
for three simulations at a density of $10^9 \gcc$ plotted as functions
of the time step number for each simulation. 
Plotting the RMS deviations against the time step number is equivalent to 
scaling the evolution time of the simulations by the fraction of 
the sound crossing time of each simulation and eliminates the differences 
in time step size due to the different resolutions. In this case
the input flame speed was $3.89 \times 10^6 \cms$.

\begin{figure}
\plotone{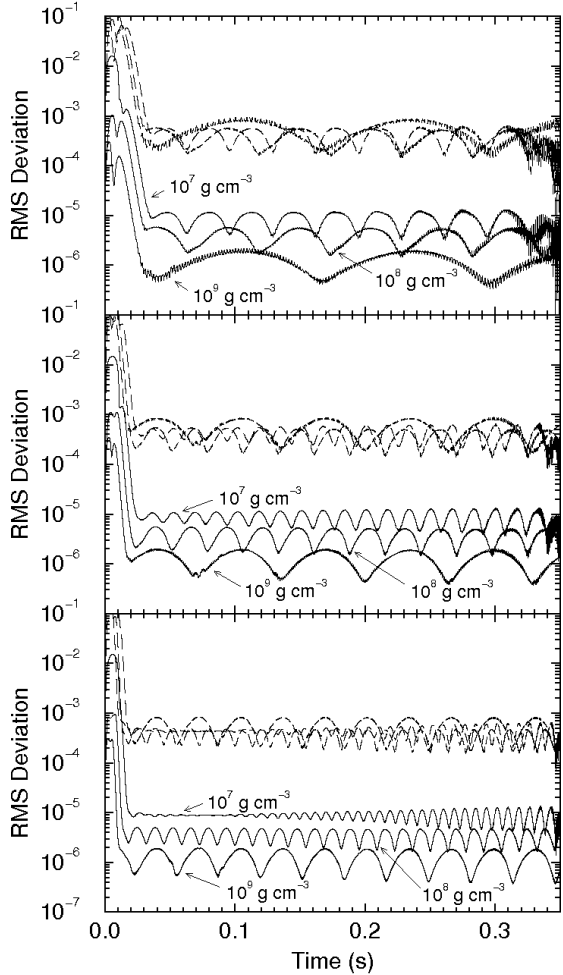}
\caption{\label{fig:pvfsp}
Plot of RMS deviations in pressure (solid curves) and 
velocity (dashed curves) for simulations of 
propagating model flames performed 
at densities of $10^7$ $10^8$, and $10^9 \gcc$ 
with a fixed
flame speed of $6 \times 10^6 \cms$. The simulations 
were performed on a uniform domain with 256 zones (top panel), 512 zones 
(middle panel), and 1024 (bottom panel) zones.
}
\end{figure}

Figure~\ref{fig:pvfsp} presents the RMS velocity and pressure deviations 
from fixed flame speed simulations performed 
at $\rho = 10^7, 10^8, 10^9 \gcc$. The panels present a set of 
simulations performed on a uniform simulation domain of 256 zones (top), 512
zones (middle), and 1024 zones (bottom).  In this figure, the RMS 
deviations are plotted against the simulation time.

\begin{deluxetable}{ccc}
\tabletypesize{\footnotesize}
\tablewidth{0pc}
\tablecaption{Simulation Properties\label{tab:noise}}
\tablehead{ \colhead{density} & \colhead{${\rm c_s}$ (fuel)} & \colhead{$t_{\rm sound crossing}$}  \\
 \colhead{$10^9\gcc$} & \colhead{$10^8\cms$} & \colhead{$10^{-3} {\mathrm{s}} $}}
\startdata
2    & 9.04 & 4.42   \\
1    & 8.06 & 4.96   \\
0.5  & 7.17 & 5.58   \\
0.3  & 6.58 & 6.08   \\
0.1  & 5.50 & 7.27   \\
0.05  & 4.82 & 8.30  \\
0.03  & 4.40 & 9.09  \\
0.01  & 3.59 & 11.1  \\
\enddata
\end{deluxetable}

\subsection{Sources of acoustic noise}
\label{sec:noisesource}

The simulations performed for this study indicate that
there are two principal sources of noise, transient noise
resulting from the initial conditions and steady rhythmic
noise produced as the model flame propagates across the
simulation domain. We observed noise from these sources 
in the simulation results in three forms, described below, one from the
initial transient and two from the propagation noise.
Though not shown here in detail, by comparing these metrics for the
multistage flame to a single-stage flame with comparable static energy
release, we confirmed that the multistage scheme adds no significant noise.

The most obvious feature of the simulations
is a large amount of noise early in the simulations. This
may be observed in all of the RMS deviation figures as the large 
deviations on the left-hand side (early time region) of the plots. 
This transient results from the method of initiating the burning.
The burned region is created by setting $Y_{{\rm ion},f},$ and $\bar q_f$ to
the appropriate values for the given density and fuel composition and the
profile of the flame to
\begin{equation}
\label{eq:initprofile}
\phi_1(x) = \frac{1}{2}\left[1.0 - \mathrm{tanh}\left[(x-x_0)/L\right]\right]
\end{equation}
where $L=b\Delta x/2$~\citep{vladimirova+05}, $\phi_2=\phi_1$, $\phi_3=0$, and
here $x_0$ is the initial position of the flame front, 5\% of the distance
across the domain.
This prescription produces initial conditions that depart
slightly from the relaxed result obtained after some evolution,
and the relaxation occurring during the initial evolution
produces the large amounts of noise observed early on.
The perturbation in this case is a large sound pulse
that propagates across the domain. 

Observation of the RMS deviation curves for a particular density
in Figure~\ref{fig:pvfsp} indicates
that the duration of the transient noise depends on 
resolution of the simulation. This occurs because the width of the 
thickened flame is set by the resolution of the simulation grid, and the
duration of this pulse is set by the flame self-crossing time. 
The wider flames at lower resolutions produce an initial transient
sound wave that has the correspondingly longer wavelength thereby
taking the correspondingly longer time to all propagate out of the domain.
As an example we consider simulations of 
Figure~\ref{fig:r1e9pv}, performed with $\rho = 10^9 \gcc$ and $s=3.89\times
10^6$~cm~s$^{-1}$. The observed times for the transient pulse to pass
completely across 
the grid were 0.016, 0.026, and 0.047 s 
for resolutions of 1024, 512, and 256 zones, respectively. 
These times were measured by observing
the pressure wave propagate across the simulation domain and agree well 
with the duration of the transients observed in the deviations.
The times are very consistent with the flame self-crossing time, $4\Delta
x/s$, which is also the time for the flame profile to come into
equilibrium, and therefore for the burning rate to stabilize.  Thus the duration
in number of time steps, $4\Delta x/s / (0.8 \Delta x/c_s) \approx 1000$, is
similar for all three resolutions, as seen in
Figure~\ref{fig:r1e9pv}.


As the flame profile moves across the regular underlying grid, the slight
changes in the profile due to the spatial quantization lead to a small,
rhythmic variation of the burning rate.  This produces a pressure wave train
propagating out through the fuel with a specific form characterized by the
quantization and with a period determined by time for the flame profile to
shift by one zone.  This leads to the high frequency oscillations readily
observed in Figure~\ref{fig:r1e9pv} after the initial transient.  
In Figure~\ref{fig:pvfsp} this noise may be seen
as the small high-frequency (period $< 0.005$ s) features on the curves and is most obvious
in the $\rho = 10^9 \gcc$ curves of the 256 zone simulation (top
panel).
As an example, we consider the simulation at $\rho = 10^9 \gcc$
with a flame speed of $s=3.89\times 10^6 \cms$ at our highest resolution, shown
in Figure~\ref{fig:r1e9pv}.  The
wave propagating through the fuel has an average wavelength of $6.76 \times
10^{5}\cm$, which with a sound speed of $8 \times 10^8 \cms$ gives a period
of $8.45 \times 10^{-4}$ s.  The flame front propagates through 16
computational zones in an average of 0.0135 s, giving $8.43 \times 10^{-4}$s
to burn each zone.  With the fuel at rest, the flame front should move within
the computational domain at $s\rho_{\rm fuel}/\rho_{\rm ash}$ (this is higher
than the flame speed due to expansion), which gives a zone crossing time of
$8.37\times 10^{-4}$~s, using the expansion for this density from
\citet{calder+07}.  All these checks prove completely consistent.

The regular oscillating pattern (with period 0.1 s and larger) of the noise visible in 
Figure~\ref{fig:pvfsp} originates from the variation in 
size of the region of the emitted wave train over
which the deviation is being averaged.  From the discussion above this wave
train has a wavelength given by equating the sound crossing time of the
disturbance with the time for the flame to cross a single zone, $\lambda/c_s
= \Delta x / v_f$.  Since the sound field is otherwise
essentially flat, averaging over an integer number of these wavelengths will
give about the same result.  Thus we expect a regular pattern in the noise
measured at a period determined by the time for the averaging interval to
shrink by one wavelength.  The interval is shrinking at the same speed that
the flame is moving across the computational domain, given above, and
dividing the wavelength of the emitted train by this gives a period of $P =
\lambda/v_f = c_s \Delta x / (s\rho_{\rm fuel}/\rho_{\rm ash})^2$.  This relation 
reproduces the
linear dependence on resolution seen in the results, and shows that the
dependence on density enters through both the sound speed and the expansion
factor.  We have confirmed that this relation reproduces the periods 
in the figures,
e.g. $P=0.24$ s for $s=6\times 10^6\cms$, $\rho = 10^9\gcc$ and our coarsest
resolution.  This result represents two bumps in the noise figures because 
we are taking the RMS deviation, losing the sign.

Finally, we note that as the flames approached the edge of the 
simulation domain and most of the fuel on the domain had burned,
the magnitude of both the high frequency oscillations and
the regular pattern in the deviation increased (as may be observed on 
the right hand side of the curves, especially at the lowest density,
$10^7\gcc$). This increase occurs because 
what little fuel remains samples the region very near the flame, 
which is expected to have the most noise.

\section{The progression of a flame from single-point initiation}
\label{sec:progression}

It is useful to describe with some detail the progress of events involved in
the GCD mechanism, and how our simulation setup captures these events.  In
the centuries before the Ia event, when the WD has accreted enough matter to
ignite carbon burning in the center, there is an expanding convective
carbon-burning core (see e.g.  \citealt{woosleyetal+04}).  This state is
already a runaway, because the temperature at the center will continue to
monotonically rise.  Ignition occurs during this convective phase when the
local heating time $\tau_{\rm heat}\simeq c_PT/\epsilon_{\rm nuc}$, where
$c_P$ is the specific heat at constant pressure and $\epsilon_{\rm nuc}$ is
the nuclear energy deposition rate, becomes shorter than the eddy turnover
time $\tau_{\rm edd}\simeq 10$-100 seconds.  At this point the burning runs
away \emph{locally}, the \carbon\ and \oxygen\ fuel converts entirely to
Fe-peak elements and a flamelet is born.

While the rate of formation of ignition points is unclear, it is believed
that ignition of local flamelets in the core of the WD is a fairly stochastic
process \citep{woosleyetal+04, wunschwoosley+04}.  For this study we will
work under the hypothesis that the ignition conditions are rare at the time
the ignition occurs.  This means that the ignition grows from a small
($\lesssim$ 1 km) region somewhere in the first temperature scale height
($\sim 400$ km) near the center of the star, and the second ignition is long
enough after this ($\gtrsim 1$ sec) to be unimportant (see
\citealt{woosleyetal+04} for a discussion of how these scales arise.) This
picture is representative of the ignition conditions found by
\citet{hoeflich+02} in their study of the pre-runaway phase, but is
somewhat in contrast to the conclusions of some of the above work
\citep[e.g.][]{woosleyetal+04}, and is essentially the opposite hypothesis to
that taken by \citet{roepke-multi+06}.
Single-point ignition is plausible within the current uncertainties and is
quite useful for our purpose here of understanding the dynamics of a flame
bubble and characterizing our numerical methods.

\subsection{General Simulation Setup}

In order to simplify our study of single-bubble dynamics we begin our
simulation with no velocity field in the stellar core.  This is, in fact, a
poor approximation to reality because the typical outer scale velocity in
the convective core is expected to be $\sim 100$ km s$^{-1}$
\citep{kulenetal+06}, which is comparable to the laminar flame propagation
speed in this part of the star \citep{timmes+92,chamulak+07}.  This means that the
strongest initial source of perturbations on the flame surface, and therefore
seeds of the later R-T modes, is likely to be the turbulence in the
convective core flow field.  The convection field is, however, not strong
enough to destroy the flamelet once it is born.  We feel neglection of the the
convective flow field in this initial study justifiable for two reasons.
First, we would like to understand the dynamics of the bubbles and flame
surface near the core first without the additional complication of the
turbulent flow.  Second, it will be challenging to interpret the effects that
a turbulent field in two dimensions subject to the imposed axisymmetry might
produce.  Any off-axis feature acts effectively as a ring, an effect that
causes enough difficulty even with static initial conditions.
Also \citet{livne+05} find that the general outcome of off-center ignitions
is not strongly effected by the presence of a convection field.

We perform two-dimensional axisymmetric simulations with the FLASH
adaptive-mesh hydrodynamics code \citep{fryxell+00,calder+02}.
We begin our simulations with at 1.38 $M_\odot$ WD with a uniform composition
of equal parts by mass of \carbon\ and \oxygen.  This model has a central
density of $2.2\times 10^9$ g cm$^{-3}$, a uniform temperature of $4\times
10^7$ K, and a radius of approximately 2,000 km.  A spherical region on the
symmetry axis is converted to burned material with $\phi_1$ given by eq.
(\ref{eq:initprofile}) with $x=|\vec r - \vec r_{\rm off}|$ and $x_0=r_{\rm bub}$,
$\phi_2=\phi_1$, and $\phi_3=0$, where $\vec r_{\rm off}$ is the location of
the center of the ignition point.  The density is chosen to maintain pressure
equilibrium with the surrounding material.  Thus the radius of the flame
bubble, $r_{\rm bub}$, is the approximate location of the $\phi_1=0.5$
isosurface, and all simulations in this paper begin with a spherical bubble
of radius 16 km.  This is the smallest bubble that is reasonably well
resolved (having $\phi_1$ very close to 1 at the center) at 4 km resolution,
that used in the parameter study presented in section \ref{sec:discussion}.
The basic parameters and some results are listed in Table \ref{tab:sims}, and
will be discussed below.

Our adaptive mesh refinement has been chosen to capture the relevant physical
features of the burning and flow at reasonable computational expense.
We choose to refine on strong density gradients everywhere, and strong
velocity gradients in the burned material.  At the beginning of the
simulation all of the star is resolved to 16 km resolution regardless of the
maximum allowed resolution for reasons of hydrostatic stability
\citep{plewa+07}.  In regions with $\rho < 5\times 10^5$ g cm$^{-3}$
refinement is not requested, which includes all of the region outside the
star.  Regions where the flame is actively propagating ($0.1 <\phi < 0.9$,
and non-trivial flame speed) are required
to be fully refined in order to properly propagate the flame front.  In the
interest of limiting computational expense, the refinement is limited to a
finest resolution of 32 km outside a radius of 2500 km, which is above the
the surface of the star in all cases treated here.  This constraint will be
relaxed in future work, but the bulk kinetic motion of the surface flow is
not expected to be affected by this choice.

\begin{figure}
\plotone{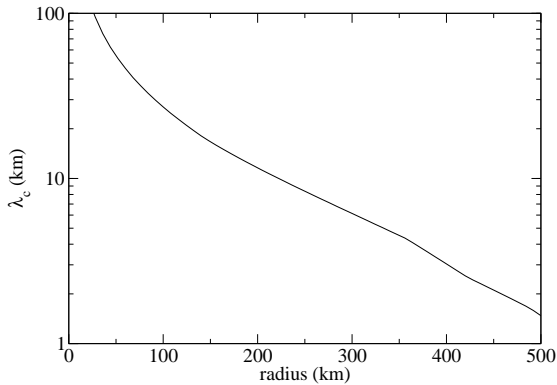}
\caption{\label{fig:lambdac}
Critical wavelength for flame surface perturbations $\lambda_c =
6\pi s^2/Ag$ in the initial WD.}
\end{figure}

\subsection{Stages of Flamelet Evolution}

The major stages in the evolution of the bubble can be roughly understood by
comparing the bubble's size to the critical wavelength for the unstable
Rayleigh-Taylor growth of flame surface perturbations.  We define
$\lambda_c \equiv 6\pi s^2/Ag$ \citep{khokhlov+95}, where $g$ is the local
gravity, $s$ is the laminar flame speed, and $A=(\rho_{\rm fuel}-\rho_{\rm
ash})/(\rho_{\rm fuel}+\rho_{\rm ash})$ is the Atwood number.  Below
$\lambda_c$, $s$ is high enough that perturbations in the flame surface can
be ``polished out'' by burning, so that the surface remains laminar and
simple.  Above $\lambda_c$, R-T is strong enough that perturbations will grow
instead, leading to a complex, disordered flame structure down to a scale of
order $\lambda_c$. (See \citealt{khokhlov+95} for an extensive discussion.)
As shown in Figure~\ref{fig:lambdac}, $\lambda_c$ drops quickly with radius,
mainly due to the increasing gravity as one moves away from the center of the
star, and after this due to the falling flame speed.
We take $r_{\rm bub} <\lambda_c$ in all cases, so that our assumption of a
spherical bubble at rest is physically justifiable \citep{fisher+07}.

\begin{figure*}
\plotone{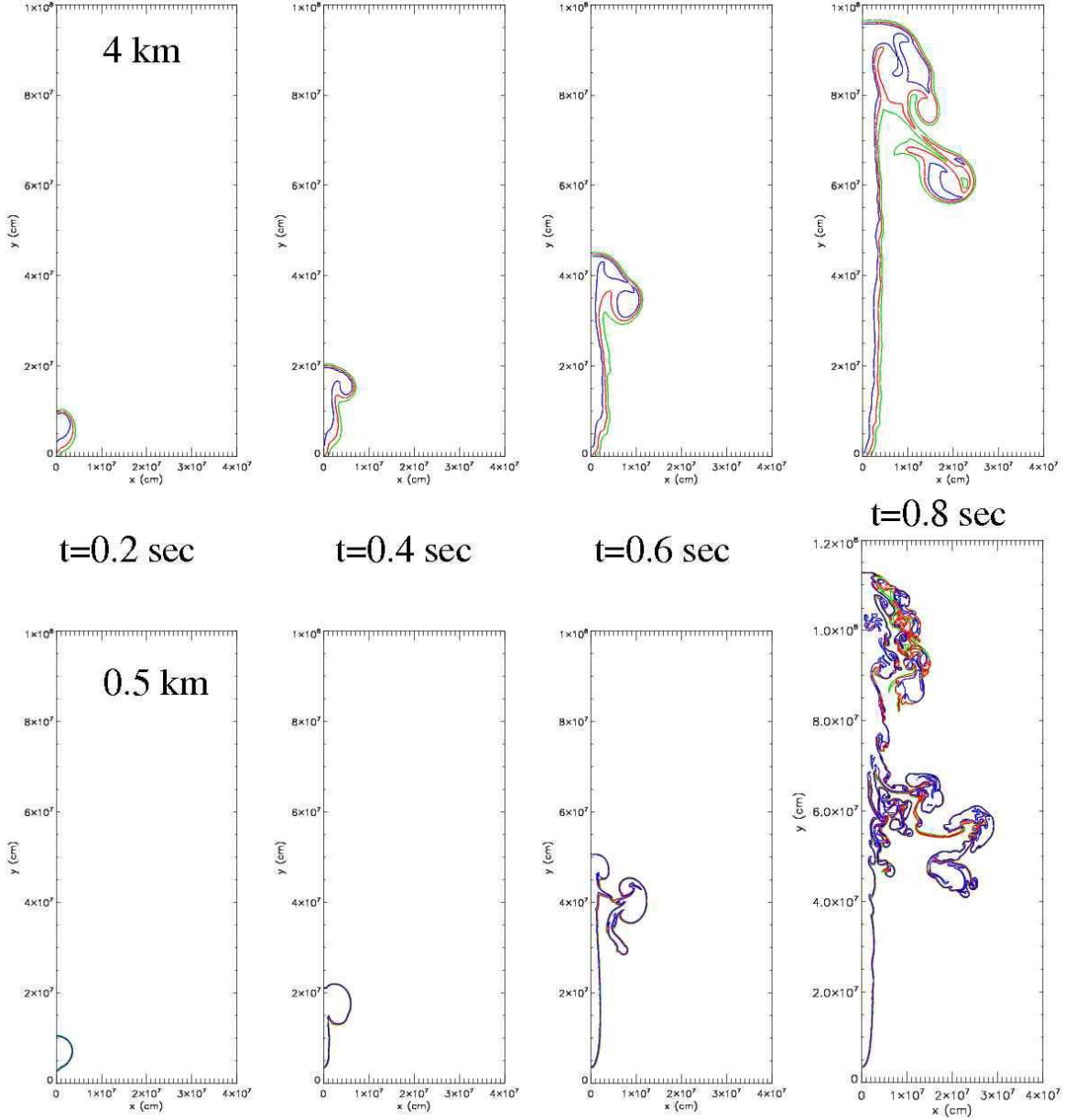}
\caption{\label{fig:bubble}
Stages of bubble growth at different resolutions.  The initial radius of the
bubble was 16 km centered at 40 km from the center of the star, contours are
shown for the progress variable $\phi=0.1$ (green), 0.5 (red), and 0.9
(blue).
}
\end{figure*}

Initially bubbles can be thought of as moving from lower left to upper right
(growing and rising) in Figure~\ref{fig:lambdac} (see \citealt{zingale+07}).
We distinguish three
phases of the bubble evolution and rise in terms of $\lambda_c$.  Each of
these phases can be seen in Figure~\ref{fig:bubble}, where we show the
evolution of the flame surface with time for two different resolutions: our
typical resolution, 4 km, and our highest resolution, 0.5 km.  Initially,
while small and near the center with $r_{\rm bub} < \lambda_c$, the flame
bubble will grow according to the laminar flame speed, roughly keeping a
spherical shape.  This structure is evident at $t=0.2$ seconds, where the
bubble has already grown to a radius of about 30 km, twice its original size.

Eventually as the bubble rises and grows, it will reach the point when
$r_{\rm bub}\sim \lambda_c$.  For our case this occurs at a little over 100
km radius.  As seen at $t=0.4$ sec, the bubble forms a R-T roll at
approximately its full dimensions, the first scale that is unstable.
This feature shows some differences with resolution, but
generally the same kind of feature (a roll) has appeared in both simulations.
Crossing $\lambda_c$, the bubble is now a R-T unstable flamelet.  The
subsequent evolution is resolved in our simulation until $\lambda_c$ falls
below the grid scale.  Thus we will term the second stage of evolution as the
resolved R-T stage.  The R-T structure becomes unresolved at approximately
350 and 500 km radius for 4 and 0.5 km resolution respectively.  Thus by
t=0.6 seconds, the 4 km simulation has already entered the unresolved regime,
while the 0.5 km simulation has nearly entered it.  This is evidenced by the
separation between consecutive contours in the progress variable caused by
strong advection of material within the flame front.

The rest of the bubble rise and burn is dominated by unresolved R-T burning.
The bubble continues to grow, both due to burning and expansion of the
material under decreasing pressure.  Its top generally reaches the surface at
approximately 1.0 seconds (``breakout''), after which it mostly exits the interior of the
star and expands strongly to create the flow around the stellar surface.
A notable difference between our evolution and that seen by
\citep{roepke-astroph+06}
is the presence of the ``stem'' below the rising bubble.  This stem is absent
in non-reactive rising bubbles \citep{Vlad07}, but it is
mysterious that it is completely absent in the level-set reactive flow
simulations.  This might be related to the fact that the initial bubbles
used by \citet{roepke-astroph+06} are too large and far off-center to capture
either the laminar or resolved R-T evolution stages, but more investigation
is necessary.  Though both simulations in Figure~\ref{fig:bubble} are
unresolved at $t=0.8$ seconds, comparing the two provides a good demonstration
of the enhancement in flame surface facilitated by smaller scale structure.

As mentioned in section \ref{sec:model}, any flame tracking method that
depends on the advection of a scalar field faces difficulties in the highly
turbulent flow produced in the unresolved R-T stage.  Measures must be taken
to compensate for the finite resolution of the simulation, otherwise the
flame ceases to propagate correctly because the turbulence destroys the
necessary structure for self-propagation of the scalar field.
This problem is particularly relevant to the simulation of WD deflagrations,
as has been extensively discussed by previous authors
\citep{khokhlov+95,gamezo+03,schmidt+06}.  R-T can create turbulence down to
the scale $\lambda_c$, below which the flame is able to polish out the mixing
perturbations.  It is thought \citep{khokhlov+95} that the details of this
small-scale behavior need not be followed explicitly.  If the overall
dynamics of the flame surface are determined by the large scale
perturbations, simulations of moderate resolution (which may however be only
just possible today in 3-d) can determine the evolution of the burning front.
This is called self-regulation, and the open question is: What resolution is
``enough''?  This is currently under debate, and we will argue here that
we have enough resolution for some purposes and can make statements about
others based on trends that we see with resolution.

\citet{schmidt+06} attempted to compensate for the shortcoming of the limited
resolution by artificially enhancing the propagation speed of the flame front
based on a measure of local turbulence.  Such complex measures should not be
necessary if self-regulation holds, thus we have taken what we consider to be
a conservative approach, making the smallest possible adjustment to the front
propagation speed and evaluating the inaccuracy directly via resolution
study.  We enforce a minimum flame speed (which therefore acts effectively as
an enhancement) that is intended to ensure $\tau_{\rm R-T} \gtrsim
\tau_{\rm flame}$ where, for flame width $\delta$, $\tau_{\rm R-T} \approx
\sqrt{\delta/Ag}$ is the characteristic R-T rise time and $\tau_{\rm
flame}\approx \delta/s$ is the flame self-crossing time.  Since our
flame width is proportional to the resolution, this leads to a limit $s \ge
\alpha\sqrt{Agm_f\Delta x}$, where $\Delta x$ is the resolution, $\alpha$ is a
geometrical factor that is different in two and three dimensions, and $m_f$ is a
calibrated constant.

We have determined $m_f$, the constant that determines when the flame model
``breaks'', empirically by running two-dimensional flame-in-a-box simulations like those
of \citet{khokhlov+95} and \citet{zhang+06}.  We evaluated the range of $s$
over which self-regulation holds, in which the burning rate, $\dot m_{\rm
burned} = \rho L^{(d-1)} s_{\rm eff}$ is determined by the box size, $L$,
such that $s_{\rm eff} = \alpha\sqrt{AgL}$, independent of $s$. In two 
dimensions $\alpha=0.28$ and in three dimensions, as found by 
previous authors, $\alpha=0.5$.  We
found that the self-regulated regime is bounded above by $s\simeq \sqrt{Ag
L}$, corresponding to the requirement
$\lambda_c\lesssim L$, and below by $s \simeq \alpha\sqrt{Ag0.04\Delta x}$.  In
our WD simulations we have used the value for $\alpha$ from three-dimensional
simulations, although we are
working in two-dimensional cylindrical geometry, and $m_f=0.04$. These values 
have been confirmed by
preliminary three-dimensional flame-in-a-box calculations, which will be
discussed in a
separate work.  Using this type of floor on $s$ requires that we explicitly
turn off the flame at low density.  This is done smoothly between
$\rho = 10^7\gcc$ and $5\times 10^6\gcc$, so that $s=0$ for densities below this.

\subsection{Resolution Study}
\label{sec:refine}

Some properties of the off-center deflagration model that we are trying to
deduce from our simulations show dependence on the simulation resolution,
while others do not.  We would like to make statements as much as
possible based on features that are not influenced by resolution, and where we
cannot avoid it, account for the dependence in other conclusions that we
draw.  Problems with resolution-dependence is not entirely unexpected, since,
as just discussed, a significant amount of our simulation is \emph{a priori}
known to be unresolved.  In summary, we find that the conditions at the
possible detonation point are fairly insensitive to resolution, for the
resolutions considered, but that the state of the interior of the star at a
given time during the runaway may only be calculated by higher resolution
simulations than the 4 km at which our parameter study was performed.

\begin{figure}
\plotone{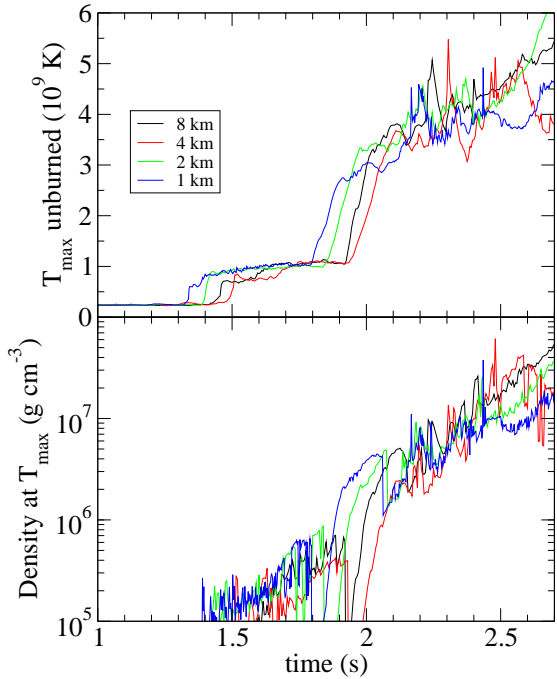}
\caption{\label{fig:tmax-rho_res}
Maximum temperature $T_{\rm max}$ in the lower hemisphere and the density at
the same point.  Shown are results from simulations where the deflagration is
ignited by a bubble of flame with a radius of 16km placed 40 km off-center
and for which the resolution is varied to 8, 4, 2, and 1 km.
Material flowing over the surface enters the lower hemisphere at
approximately 1.5 seconds and the collision occurs at approximately 2.0
seconds, at which point the material near the collision region begins to
compress.  We find that the conditions at the ignition point are insensitive
to the simulation resolution.
}
\end{figure}

Of foremost interest is the robustness of the gravitationally confined
detonation (GCD) mechanism, particularly the properties in the collision and
compression regions opposite the eruption point.  We have
performed two-dimensional simulations of various resolutions that begin from a 16 km
radius burned region offset from the center by 40 km.  The history of the
maximum temperature, $T_{\rm max}$, in the lower hemisphere ($\theta >
\pi/2$), and the density
at that same point is shown in Figure~\ref{fig:tmax-rho_res}.  After the
bubble has reached the surface, material -- burned and unburned -- begins to
flow over the surface of the star towards the opposite pole.
This material enters the lower hemisphere
approximately 1.5 seconds after the ignition.  Some of the
low-density material is being shocked as it interacts with the stellar
surface as it is moving around the star, giving a temperature near $10^9$ K.  The collision of material at
the lower pole occurs at approximately 2 seconds and the density of the
hottest region steadily increases thereafter until the expected ignition of
the detonation at $T>3\times 10^9$ K and $\rho >10^7\gcc$.  See section \ref{sec:discussion} for a more extensive
discussion of the fluid motions in the simulation.  We find that the
temperature and density reached at the possible ignition point
of the detonation are insensitive to the simulation resolution.  This gives
confidence in the robustness of the GCD scenario as a whole, in that the bulk
motion of the surface flow seems to be insensitive to resolution.
Sensitivity to initial conditions and assumed symmetry have not been
addressed here, and will be the subject of future work.

\begin{figure}
\plotone{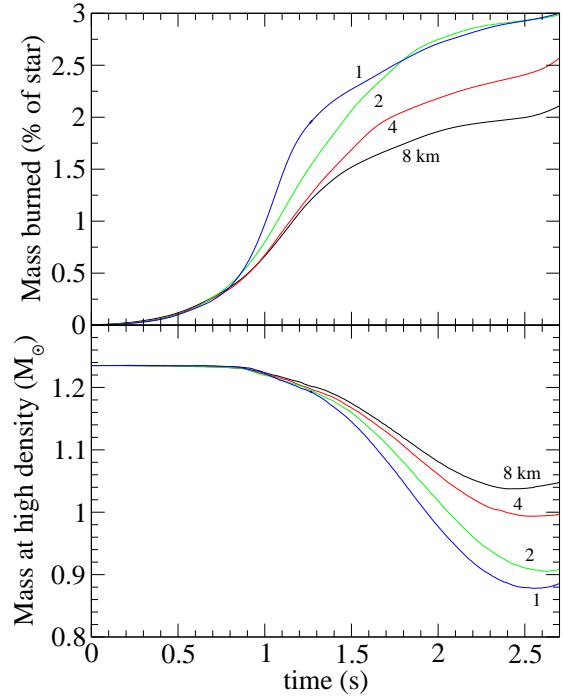}
\caption{\label{fig:mburn-mni_res}
Burned mass (as a fraction of the star) and stellar mass with $\rho > 5.5\times
10^7$ g cm$^{-3}$ for simulations shown in figure \ref{fig:tmax-rho_res}.
There is a clear dependence on resolution, with finer
resolutions generally leading to more rapid evolution.
}
\end{figure}

The overall amount of material burned, and therefore the amount of energy
added to the star, is important for determination of the interior state
(notably density) when the detonation wave sweeps through.  This will
set the amount and distribution of \nickel\ and intermediate mass
elements, and is therefore extremely important for the predictive power of
our simulations.  Shown in Figure~\ref{fig:mburn-mni_res} is the amount of
mass burned by the rising flame and the mass of the star, which has $\rho >
5.5\times 10^7$ (see next section).  As seen in Figure~\ref{fig:bubble}, it
appears that the bubble rises somewhat faster at higher resolutions, possibly
due to the decreased numerical viscosity or to faster burning during the resolved R-T
stage.  Both the total burned mass and the amount of high density material
show significant dependence on resolution, but it appears that convergence in
the overall quantities may be just within reach.  The total burned mass at 2
km and 1 km resolution is quite consistent by the time the detonation might occur, and
the difference between the amount of high density material is also fairly
modest considering the factor of 2 difference in resolution and the expected
offset in time due to the lower numerical viscosity.

But not all the news is good.  Performing a full suite of simulations
at 1 km was not undertaken for this study and will be prohibitive in three
dimensions, but knowing how far away convergence is lends great interpretive
power to our lower resolution simulations.  There is also a caution to be
raised.  Figure~\ref{fig:mburn-mni_res} does not include results at 0.5 km,
although the beginning of the curve is obviously available, because much more
mass is burned in this case.  This occurs because while the dominant trailing
vortices follow the main bubble out of the star at all lower resolutions, at
0.5 km the branch feature visible in Figure~\ref{fig:bubble} at a radius 
of 600 km does not.
This piece of flame, which has now become a ring due to the axisymmetry,
stays at high density and continues to burn a significant portion of the
star.  It is unfortunately not possible for us to judge whether this is
realistic or largely an artifact of the very different nature of
vorticity conservation in two dimensions.  We do note that simulations with a
smaller bubble (2~km) and slightly larger offset (60~km) do
not show this anomaly, and initially proceed much as those at lower
resolution.  This effect deserves close scrutiny as more simulations are
performed, especially in 3-d.  Also the shed vortices would likely not have
an adverse impact if we were simulating a centrally ignited deflagration that
consumed most of the WD on the way to the surface.

\begin{deluxetable*}{cccccccc}
\tabletypesize{\footnotesize}
\tablecolumns{8}
\tablewidth{0pc}
\tablecaption{Properties at Detonation Ignition\label{tab:sims}}
\tablehead{ \colhead{$r_{\rm bub}$} & \colhead{$r_{\rm off}$} &
\colhead{resolution} &
\colhead{$t_{\rm det}$\tablenotemark{a}} &
\colhead{$M_{\rm burn}$} &
\colhead{mass at high density} &
\colhead{max density} &
\colhead{energy release}
\\
\colhead{(km)} & \colhead{(km)} & \colhead{(km)} &
\colhead{(s)} & \colhead{(\% of star)} &
\colhead{($M_\odot$)} &
\colhead{($10^8\gcc$)} &
\colhead{($10^{50}$ erg)}
}
\startdata
\cutinhead{Resolution Study}
16 & 40 & 8 & 2.35 & 1.97 & 1.04 & 7.6 & 0.329\\
16 & 40 & 4 & 2.31 & 2.33 & 1.01 & 6.6 & 0.388\\
16 & 40 & 2 & 2.37 & 2.90 & 0.926 & 5.0 & 0.478\\
16 & 40 & 1 & 2.35\tablenotemark{b} & 2.88 & 0.892 & 4.5 & 0.517\\
\cutinhead{Offset Study}
16 &100 & 4 & 2.02 & 1.13 & 1.13 & 12 & 0.193\\
16 & 80 & 4 & 2.05 & 1.36 & 1.12 & 11 & 0.226\\
16 & 50 & 4 & 2.19 & 2.07 & 1.05 & 8.0 & 0.347\\
16 & 40 & 4 & 2.31 & 2.33 & 1.01 & 6.6 & 0.388\\
16 & 20 & 4 & 2.70\tablenotemark{c} & 6.57 & 0.473 & 1.6 & 1.15
\enddata
\tablenotetext{a}{$t_{\rm det}$ is defined as the first time at which $\rho
> 10^7\gcc$ at the point of $T_{\rm max} > 3\times 10^9$ K.}
\tablenotetext{b}{In this case we have neglected the early, short-lived,
fluctuation at $t\simeq 2.15$ s}
\tablenotetext{c}{Our conservative detonation criteria are not reached,
listed are values for the peak density of the $T_{\rm max}$ point.}
\tablecomments{All values are evaluated at the time indicated, $t_{\rm
det}$.}
\end{deluxetable*}

\section{The distributions of outcomes prior to possible detonation}
\label{sec:discussion}

The uncertainty in the location of the initial flamelet, discussed in section
\ref{sec:progression},
leads us to consider ignition of a flame bubble at several offsets,
$r_{\rm off}$, from the center of the star.  We find that both the time at
which the detonation conditions are reached at the point opposite bubble
eruption and the expansion of the star up to this time are correlated with
$r_{\rm off}$.  Since the expansion of the star is directly related to
the density of the material through which the detonation will propagate, the
observed result will be a variation in the \nickel\ mass ejected, and that of
intermediate elements.

\begin{figure}
\plotone{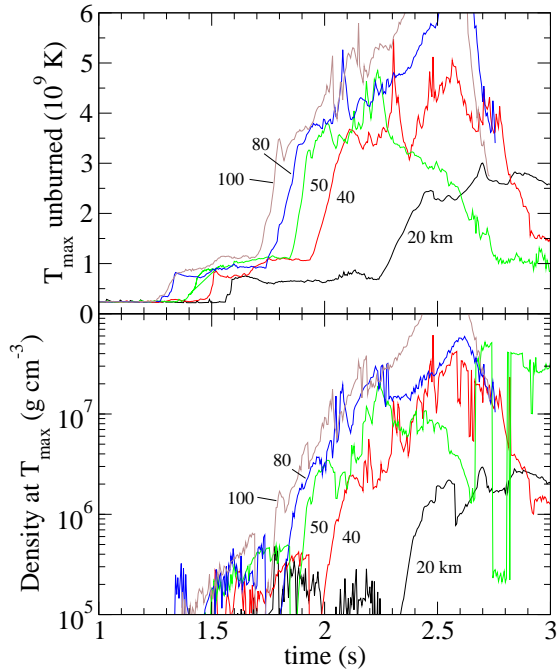}
\caption{\label{fig:tmax-rho_param}
Maximum temperature $T_{\rm max}$ in the lower hemisphere and the density at
the same point.  Shown are results from simulations at 4 km resolution with
an ignited bubble of 16 km radius placed at several offsets from the center
of the star, 20, 40, 50, 80, and 100 km.
Material flowing over the surface enters the lower hemisphere at
approximately 1.5 seconds and the collision occurs at approximately 2.0
seconds, varying some with offset, at which point the material near the
collision point begins to compress.
}
\end{figure}

The maximum temperature, $T_{\rm max}$, in the lower hemisphere and the density at the same
point are shown in Figure~\ref{fig:tmax-rho_param} for a range of offsets from
20 to 100 km.  The rise in temperature near $t=1.5$ seconds is when the
material flowing over the surface of the star passes the equator. Collision
of the surface flow at the lower pole occurs at a variety of times between
1.7 and 2.4 seconds, where $T_{\rm max}$ rises to several $10^9$ K, and the
density steadily increases.   Conservative detonation conditions require
$T\gtrsim 3\times 10^9$ K and $\rho \gtrsim 10^7$ g cm$^{-3}$
\citep{roepke-astroph+06,niemeyer+97}.  These are met
at 2.02, 2.05, 2.19 and 2.31 seconds for 100, 80, 50 and 40 km respectively.
Several other properties of the star at the time when the ignition is
expected to occur are listed in Table \ref{tab:sims}, particularly the total
energy released up to that point, in addition to the total burned mass.
Note that the total binding energy is $4.95\times 10^{51}$ erg, so that none of
cases here come close to unbinding the star; that is expected to occur during
the detonation phase.
Having only flame burning, our models continue after the detonation point and
show that the detonation conditions appear to be robust and long-lived,
especially at the larger offsets.  Our 20 km simulation expands the star much
more by the time of the collision, due to more burning occurring in the
interior, and may not reach conditions sufficient for detonation.  This
certainly indicates that the GCD mechanism is likely to fail for ignition
points very close to the center.

\begin{figure}
\plotone{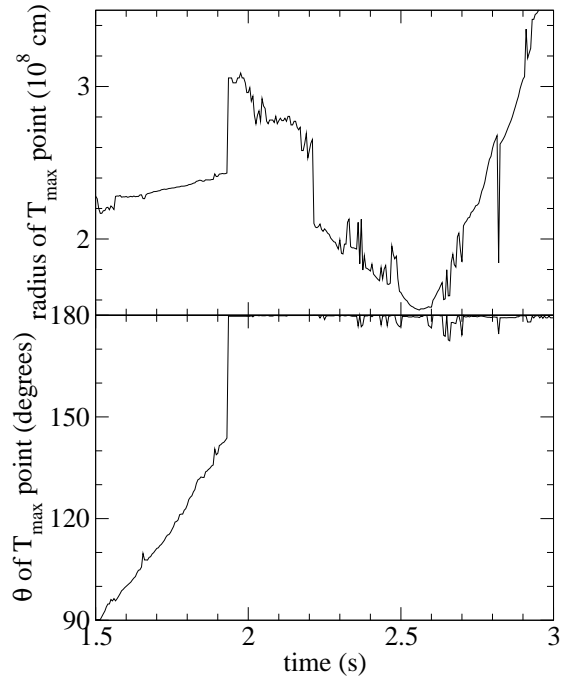}
\caption{\label{fig:r-theta}
Radius and polar angle, $\theta$, location of $T_{\rm max}$ point for
simulation with $r_{\rm bub}=16$ km and $r_{\rm off}= 40$ km.  The region
above the surface ($r\simeq 2\times 10^8 \cm$) is heated as material
collides.  Material pushed toward the star from the collision point then
interacts with the stellar surface to form a hot, dense region that
penetrates into the outer layers of the star.
}
\end{figure}

The radius and the polar angle ($\theta$) of the maximum temperature point
are shown in Figure~\ref{fig:r-theta}, demonstrating the evolution of
conditions that lead to the detonation ignition.
The initial surface of the star is
at $r=2\times 10^8$ cm.  Thus we see that the temperature maximum moves
around the surface between 1.5 and 2.0 seconds, then shifts to the pole at the
collision.  Note that this jump is not material motion.  At approximately 2.2
seconds,
material confined between the
collision point and the star becomes the hottest, and the hot spot moves
steadily inward from $2\times 10^8$ cm.  During this time the hot spot is not
always precisely on the axis.  Eventually the compression subsides and this
hot spot dissipates.
While the hot spot moves into the star with a speed of approximately
$10^8$~cm~s$^{-1}$, material ahead of it (closer to the star) is nearly
at rest, and that behind it is moving in at just above $10^9$~cm~s$^{-1}$, so
that the hot spot occurs in the accumulation.

\begin{figure}
\plotone{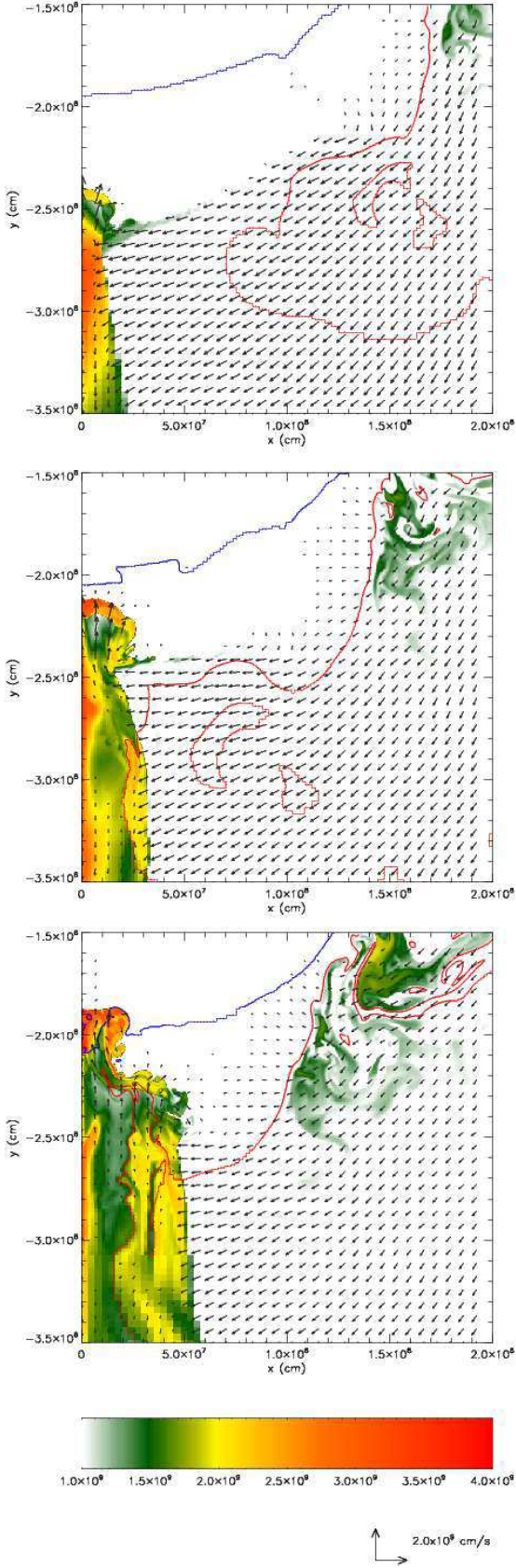}
\caption{\label{fig:collision}
Detail of flow near the collision and detonation region for the $r_{\rm off}
= 40$~km case, from top to bottom at $t=2.07$, 2.19, and 2.32 seconds.
Temperature is shown in color and contours are shown at
$\rho=10^7$~g~cm$^{-3}$ (blue) and at the edge of the burned material ($\phi =
0.1$, red).  Velocity vectors smaller that $10^8$~cm~s$^{-1}$ are not shown.
A stagnation point is formed above the surface of the star
from which material is projected out along the axis and compressed against the
surface of the star, where the detonation is expected to occur.
}
\end{figure}

A detailed view of the flow near the collision region is shown in 
Figure~\ref{fig:collision}, where we see that a stagnation point is formed in the
colliding unburned material.  From this, material is projected out along the
axis and toward the stellar surface, compressing the surface layers of the
star toward ignition conditions.  The phrase ``gravitational confinement''
does not convey the full impression of what is occurring. The detonating
material appears to be inertially confined by flow originating at the
collision point, although the amount of compression occurring likely reflects
both the strongly gravitationally stratified WD surface and a certain amount
of assistance from gravity, such that both high gravity and and a flow with
significant inertia are required to reach such high temperatures and
densities.  The collision itself arises because the material is
gravitationally bound, however, it is the kinetic motion imparted to the
material by the expanded bubble at the breakout point that eventually leads
to the (gravitationally assisted) confinement.

\begin{figure}
\plotone{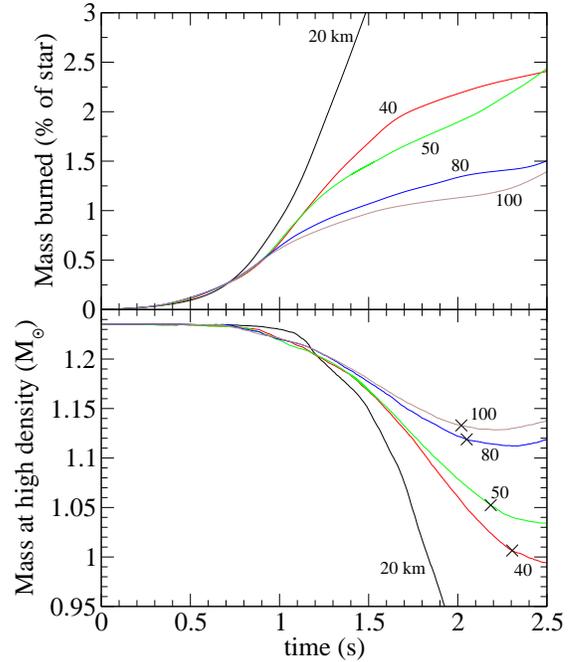}
\caption{\label{fig:mburn-mni_param}
Burned mass (as a fraction of the star) and stellar mass with $\rho > 5.5\times
10^7$ g cm$^{-3}$ for same simulations as in Figure~\ref{fig:tmax-rho_param}.
The time at which a detonation is expected to be launched is marked
with a $\times$ for each case.  Larger offsets are expected to produce more
\nickel\ in the ejected material.
}
\end{figure}

In the GCD scenario, because so little material is burned during the
deflagration phase, the amount of \nickel\ produced in the supernova is
determined by the density distribution during the detonation phase.  In lieu
of simulating the propagation of the detonation, which will be performed in
future work, we have measured the mass of material above $5.5\times 10^7$ g
cm$^{-3}$.  This limit is obtained from the density at which material in the W7
model \citep{nomoto+84} burned to only 50\% \nickel.  This is obviously only a
rough estimate, but is good enough for measuring the trend with offset
distance that we are interested in here.  The bottom panel of 
Figure~\ref{fig:mburn-mni_param} shows how this possible \nickel\ mass decreases as
the star expands during the deflagration phase.  The curves are marked at the
expected launch time of the detonation, where the temperature and density
first exceed $3\times 10^9$ K and $10^7$ g cm$^{-3}$ together.  

We find that the amount of \nickel\ expected in the ejecta is correlated with
the offset of the initial (small) ignition region.  Larger offsets can
produce more \nickel\ for two reasons: (1) less energy is released in the
deflagration phase, and therefore the star has expanded less when the
detonation occurs, and (2) the detonation conditions happen sooner so that
the star has had less time to expand.  It does appear that the first of these
is the dominant effect.  The top panel of Figure~\ref{fig:mburn-mni_param}
shows the mass burned as a fraction of the star with time.  Larger offsets
burn less of the star during the bubble rise and breakout, leading to less
expansion of the star.

The \nickel\ mass estimates we have found here are fairly high, but as seen
in section \ref{sec:refine} this resolution (4 km) appears to somewhat
underestimate the burned mass and overestimate the possible \nickel\ at the
detonation time.  We are not claiming to have performed an absolute
calculation of the \nickel\ mass for a given ignition point offset; we have
instead demonstrated a trend that appears to be robust with respect to the
physical processes that are occurring.  We hope that in the future, with
higher resolution such that self-regulation of the burning is strong enough
that we can constrain the R-T phase better, we may be able to construct a
fully predictive model.  There are, however several steps that should be
taken in the mean time, including three-dimensional studies that are underway
\citep{jordan+07}, and studies
of flame bubble response to the strong convection expected to be present in
the WD core when the ignition occurs.  The current level of calculation is,
however sufficient for measuring trends such as how things might change with
the relative C/O fraction in the interior of the WD.

\section{Conclusions}
\label{sec:conclusion}

We have shown that in the GCD picture of a delayed detonation of a WD
near the Chandrasekhar mass, the properties of the WD at detonation, notable the
density distribution, are systematically correlated with the offset of the
ignition point of the deflagration.  Assuming that the detonation phase
proceeds as in previous simulations, this will cause a variation in the
\nickel\ mass ejected in the supernova.  The position of the ignition point
within the inner few 100 km of the WD is expected to be stochastically
determined by the turbulent flow in this region.  GCD thus provides a
possible explanation for the variety of \nickel\ masses seen in Type Ia
Supernovae.

We find that the conditions (temperature and density reached) at the
candidate launch point of the detonation are insensitive to the resolution of
the simulation for resolutions studied here ($\le 8$~km).  This is a good
mark for the robustness of the GCD mechanism, but more work is needed,
especially related to the possibility of vortex shedding early in the bubble
rise and the strong convection that should be present in the core at the
time of ignition.  We have indications of numerical convergence in both the
total burned mass and the mass of dense material, and therefore the predicted
\nickel\ mass produced by a given ignition offset.  But caution is advisable:
the mass burned during the highly Rayleigh-Taylor (buoyancy-driven) unstable rise of
the burned region through the star is seen to vary with resolution, generally
progressing faster with higher resolution, even though convergence in the
final value appears to have been reached.  Also, converged results (in the
extremely limited sense indicated here) appear to require 2 km or possibly 1
km resolution, which is prohibitive in three dimensions.  Even here, our
parameter study has been performed at 4 km resolution for efficiency.  Thus
we are able to predict trends in the \nickel\ mass, but not the actual value
ejected for a given offset.

Our method for following the nuclear energy release, including neutronization,
with an ADR flame model was described in detail.  This method reproduces the
energy release and hydrodynamic characteristics of the nuclear burning by
following a limited number of parameters coupled to an artificially thickened
flame front.  We have demonstrated that the energy release adds a minimal
amount of unwanted acoustic noise (RMS velocity $<\rm few\times10^{-4}$) to the
simulation, largely removing this source of unrealistic seeds for the
instabilities in the rising flame surface.

\acknowledgments
The authors thank Alexei Khokhlov for encouragement and insight during
development of the flame model, Robert Fisher for enlightening discussions
during the later stages of this work, George Jordan for preliminary work
implementing an electron-ion formalism in Flash, the code group at the
ASC/Flash center, especially Anshu Dubey and Dan Sheeler, for support in code
development, and Ed Brown for comments on the manuscript.
This work is supported at the University of Chicago in part by the National
Science Foundation under Grant PHY 02-16783 for the Frontier Center ``Joint
Institute for Nuclear Astrophysics'' (JINA), and in part by the U.S.
Department of Energy under Contract B523820 to the ASC Alliances Center for
Astrophysical Flashes.  ACC acknowledges support from the NSF grant
AST-0507456.  JWT acknowledges support from Argonne National Laboratory,
which is operated under contract No.\ W-31-109-ENG-38 with the DOE.

\bibliographystyle{apj}

\end{document}